\documentclass[11pt]{article}
\usepackage{graphicx, amssymb}
\newcommand{\SetFigFont}[3]{}

\title{A Variational Principle in Discrete Space-Time -- \\
Existence of Minimizers}
\author{Felix Finster}
\date{March/October 2005}

\newtheorem{Def}{Def.}[section]
\newtheorem{Thm}[Def]{Theorem}
\newtheorem{Prp}[Def]{Proposition}
\newtheorem{Lemma}[Def]{Lemma}

\newtheorem{Corollary}[Def]{Corollary}
\newtheorem{Example}[Def]{Example}
\newcommand{\Proof}{{\em{Proof. }}}
\newcommand{\QED}{\ \hfill $\FBox$ \\[1em]}
\newcommand{\QEDrem}{\ \hfill $\blacklozenge$}
\newcommand{\spc}{\;\;\;\;\;\;\;\;\;\;}

\newcommand{\bra}{\mbox{$< \!\!$ \nolinebreak}}
\newcommand{\ket}{\mbox{\nolinebreak $>$}}
\newcommand{\C}{\mathbb{C}}
\newcommand{\R}{\mathbb{R}}
\newcommand{\1}{\mbox{\rm 1 \hspace{-1.05 em} 1}}

\newcommand{\N}{\mathbb{N}}
\newcommand{\sN}{\mbox{\rm \scriptsize I \hspace{-.8 em} N}}

\newcommand{\Tr}{\mbox{\rm{Tr}\/}}

\newcommand{\beq}{\begin{equation}}
\newcommand{\eeq}{\end{equation}}

\newcommand{\FBox}{\rule{2mm}{2.25mm}}

\begin{document}
\maketitle

\begin{abstract}
We formulate a variational principle for a collection of
projectors in an indefinite inner product space.
The existence of minimizers is proved in various situations.
\end{abstract}

In a recent book it was proposed to formulate physics with a new variational
principle in space-time~\cite{PFP}. In the present paper we
construct minimizers of this variational principle. In order to make the
presentation self-contained and easily accessible, we introduce the mathematical
framework from the basics (see Sections~\ref{sec1} and~\ref{sec2}). Thus this
paper can be used as an introduction to the mathematical setting of the
principle of the fermionic projector. However, the reader who wants to get
a physical understanding is referred to~\cite{PFP}.

Our variational principle is set up in finite dimension, and thus the
continuity of the action is not an issue. The difficulties are
the lack of compactness and the fact that there is no notion of convexity.
Therefore, we need to derive suitable estimates (Sections~\ref{sec4}
and~\ref{sec6}) before we can use the direct method of the calculus of
variations (Sections~\ref{sec7} and~\ref{sec8}). Our main results are
stated in Section~\ref{sec2}, whereas in Section~\ref{sec3} we explain
our variational principle and illustrate it with a few simple examples.

\section{Discrete Space-Time and the Fermionic Projector} \label{sec1}
Let~$H$ be a finite-dimensional complex vector space,
endowed with a sesquilinear form $\bra .|. \ket \::\:
H \times H \to \C$, i.e.\ for all $u, v, w \in H$ and $\alpha, \beta \in \C$,
\begin{eqnarray*}
\bra u \:|\: \alpha v + \beta w \ket &=& \alpha \:\bra u \:|\: v \ket +
\beta \:\bra u \:|\: w \ket \\
\bra \alpha u + \beta v \:|\: w \ket &=& \overline{\alpha} \:\bra u \:|\: w \ket
+ \overline{\beta} \:\bra v \:|\: w \ket\:.
\end{eqnarray*}
We assume that~$\bra .|. \ket$ is symmetric,
\[ \overline{\bra u \:|\: v \ket} \;=\;\: \bra v \:|\: u \ket \:, \]
and non-degenerate,
\[ \bra u \:|\: v \ket \;=\; 0 \;\;\; \forall \:v \in H \quad \Longrightarrow \quad
u \;=\; 0 \:. \]
Note that~$\bra .|. \ket$ is in general not positive, and it is therefore
not a scalar product. We also refer to~$(H, \bra .|. \ket)$ as an
{\em{indefinite inner product space}}. To a non-degenerate subspace of~$H$
we can associate its {\em{signature}} $(p,q)$, where~$p$ and~$q$ are the
maximal dimensions of positive and negative definite subspaces, respectively
(for more details see~\cite{Bognar, GLR} and the examples in Section~\ref{sec3}).

Many constructions familiar from scalar product spaces can be carried over to
indefinite inner product spaces. In particular, we define the {\em{adjoint}} of
a linear operator~$A \::\: H \to H$ by the relation
\[ \bra u \:|\: A v \ket \;=\; \bra A^* u \:|\: v \ket \qquad
\forall \:u, v \in H\:. \]
A linear operator~$A$ is said to be {\em{unitary}} if~$A^*=A^{-1}$
and {\em{symmetric}} if $A^*=A$. It is called a {\em{projector}} if it is
symmetric and idempotent,
\[ A^* \;=\; A \;=\; A^2\:. \]

Let~$M$ be a finite set. To every point~$x \in M$ we associate a projector
$E_x$. We assume that these projectors are orthogonal and
complete in the sense that
\beq \label{oc}
E_x\:E_y \;=\; \delta_{xy}\:E_x \spc {\mbox{and}} \spc
\sum_{x \in M} E_x \;=\; \1\:.
\eeq
Equivalently, we can say that the images of the projectors~$E_x$
give a decomposition of~$H$ into orthogonal subspaces,
\beq \label{odecomp}
H \;=\; \bigoplus_{x \in M} E_x(H)\:.
\eeq
Furthermore, we assume that the images~$E_x(H) \subset H$ of these
projectors are non-de\-ge\-ne\-rate and all have the same signature~$(n,n)$.
We refer to~$(n,n)$ as the {\em{spin dimension}}.
Relation~(\ref{odecomp}) shows that the dimension of~$H$ must be equal
to~$m \!\cdot\! 2n$,
where~$m=\# M$ denotes the number of points of~$M$.
The points~$x \in M$ are
called {\em{discrete space-time points}}, and the corresponding
projectors~$E_x$ are the {\em{space-time projectors}}. The
structure~$(H, \bra .|. \ket, (E_x)_{x \in M})$ is
called {\em{discrete space-time}}.

We now introduce one more projector~$P$ on~$H$, the so-called
{\em{fermionic projector}}, which has the additional property that its image~$P(H)$ is
{\em{negative}} definite. In other words, $P(H)$ has signature~$(0,f)$
with~$f \in \N$.
The vectors in the image of~$P$ have the interpretation as the
quantum mechanical states of the
particles of our system, and we call~$f=\dim P(H)$ the {\em{number of particles}}.
We remark that in physical applications~\cite{PFP} these particles are Dirac particles,
which are fermions, giving rise to the name ``fermionic projector''.

A space-time projector~$E_x$ can be used to restrict an operator
to the subspace~$E_x(H) \subset H$. Using a more graphic notion, we
also refer to this restriction as the {\em{localization}} at the
space-time point~$x$.
For example, using the completeness of the space-time projectors~(\ref{oc}),
we readily see that
\beq \label{loctr}
f \;=\; \Tr \,P \;=\; \sum_{x \in M} \Tr (E_x P)\:.
\eeq
The expression~$\Tr (E_x P)$ can be understood as the localization of the trace
at the space-time point~$x$, and summing over all space-time points
gives the total trace. We call $\Tr (E_x P)$ the {\em{local trace}} of~$P$.
When forming more complicated composite expressions in the
projectors~$P$ and~$(E_x)_{x \in M}$, it is convenient to use
the short notations
\[ P(x,y) \;=\; E_x\,P\,E_y \spc {\mbox{and}} \spc
u(x) \;=\; E_x\,u \:. \]
Referring to the orthogonal decomposition~(\ref{odecomp}), $P(x,y)$
maps~$E_y(H)$ to~$E_x(H)$ and vanishes otherwise. It is often useful
to regard~$P(x,y)$ as a mapping only between these subspaces,
\[ P(x,y)\;:\; E_y(H) \: \rightarrow\: E_x(H)\:. \]
Using~(\ref{oc}), we can write the product~$Pu$ as follows,
\[ (Pu)(x) \;=\; E_x\: Pu \;=\; \sum_{y \in M} E_x\,P\,E_y\:u
\;=\; \sum_{y \in M} (E_x\,P\,E_y)\:(E_y\,u) \:, \]
and thus
\[ (Pu)(x) \;=\; \sum_{y \in M} P(x,y)\: u(y)\:. \]
This relation resembles the representation of an operator with an integral kernel.
Therefore, we call~$P(x,y)$ the {\em{discrete kernel}} of the fermionic projector.
The discrete kernel can be used for expressing general operator products; for example,
\[ P\:E_x\:P\:E_y \;=\; \sum_{z \in M} P(z,x)\: P(x,y)\:. \]

\section{A Variational Principle, Statement of the Main Results} \label{sec2}
We want to form a positive quantity which depends on the form of the fermionic
projector relative to the space-time projectors. Since scalar invariants
(like the trace or the determinant) can be introduced only for operators
which map a vector space into itself, we first define the
{\em{closed chain}}~$A_{xy}$ by
\beq \label{cc}
A_{xy} \;=\; P(x,y)\: P(y,x) \;=\; E_x \:P\: E_y \:P\: E_x \;:\;
E_x(H) \: \rightarrow\: E_x(H)\:.
\eeq
We shall often omit the subscripts `xy'.
Let~$\lambda_1,\ldots,\lambda_{2n}$ be the zeros of the characteristic polynomial
of~$A$, counted with multiplicities. We define the {\em{spectral weight}}~$|A|$ by
\[ |A| \;=\; \sum_{j=1}^{2n} |\lambda_j|\:. \]
More generally, one can take the spectral weight of powers of~$A_{xy}$, and by summing
over the space-time points we get positive numbers depending only on the
projectors~$P$ and~$(E_x)_{x \in M}$.

For a given parameter~$\kappa>0$ we consider the family of fermionic
projectors~${\mathcal{P}}(\kappa)$ defined by
\beq \label{constraint}
{\mathcal{P}}(\kappa) \;=\; \Big\{ P {\mbox{ with }}
\sum_{x,y \in M} |A_{xy}|^2 = \kappa \Big\} \:.
\eeq
Our variational principle as introduced in~\cite[\S3.5]{PFP} is to
\beq \label{vary}
{\mbox{minimize}} \quad \sum_{x,y \in M} |A_{xy}^2| \quad
{\mbox{by varying~$P$ in ${\mathcal{P}}(\kappa)$}}\:,
\eeq
keeping the number of particles~$f$ as well as discrete space-time
$(H, \bra .|. \ket, (E_x)_{x \in M})$ fixed.
The next theorem completely settles the existence problem.
\begin{Thm} \label{thmn1}
For every~$\kappa$ for which the family~${\mathcal{P}}(\kappa)$ is not empty,
the variational principle~(\ref{vary}) attains its minimum.
\end{Thm}
This theorem makes no statement on uniqueness, and indeed we do not see a reason
why the minimizers should be unique. Using the method of Lagrangian
multipliers\footnote{{\textsf{Footnote added in January 2013}:
For clarity, we point out that we here apply the Lagrange multiplier method
in a naive way. Verifying that the constraints are regular is rather subtle;
for details we refer the reader to the recent article~arXiv:1205.0403 [math-ph].}},
for every minimizer~$P$ there is a real parameter~$\mu$ such that~$P$
is a stationary point of the {\em{action}}
\beq \label{Sdef}
{\mathcal{S}}_\mu[P] \;=\; \sum_{x,y \in M} {\mathcal{L}}_\mu[A_{xy}]
\eeq
with the {\em{Lagrangian}}
\beq \label{Ldef}
{\mathcal{L}}_\mu[A] \;=\; |A^2| - \mu\: |A|^2 \:.
\eeq

Unfortunately, the above theorem does not give
information on the value of the Lagrangian multiplier~$\mu$. Knowing~$\mu$
is important because in physical applications the Lagrangian multiplier is determined
by the model (more precisely, for the fermionic projector of the standard
model~\cite[Chapter~5]{PFP} one takes~$n=16$ and~$\mu=1/28$, whereas for modeling
the simpler system of one sector one takes~$n=2$ and $\mu=1/4$). Thus we would like
to construct stationary points of the action~(\ref{Sdef}) for a given value of~$\mu$.
The simplest method to achieve this is to minimize directly the action~(\ref{Sdef}) in the
class of all fermionic projectors. This is our motivation for considering also the
variational principle
\beq \label{vp}
{\mbox{minimize ${\mathcal{S}}_\mu[P]$ by varying~$P$}}\:,
\eeq
again keeping the number of particles as well as discrete space-time fixed.
We point out that every minimizer~$P$ of the variational principle~(\ref{vp}) is
also a minimizer of the variational principle with constraint~(\ref{vary},
\ref{constraint}) for the corresponding value of~$\kappa =
\sum_{x,y} |P(x,y) P(y,x)|^2$. Since we regard the variational principle~(\ref{vp})
merely as a technical tool for constructing minimizers of~(\ref{vary}), we refer
to~(\ref{vp}) as the {\em{auxiliary variational principle}}.

Let us discuss the behavior of the auxiliary variational principle for different
values of~$\mu$. Clearly, a necessary condition for the existence of minimizers is
that the action is bounded from below. According to the Schwarz inequality,
\[ |A| \;=\; \sum_{j=1}^{2n} |\lambda_j| \;\leq\;
\left( \sum_{j=1}^{2n} 1 \right)^\frac{1}{2}
\left( \sum_{j=1}^{2n} |\lambda_j|^2 \right)^\frac{1}{2}
\;=\; \sqrt{2n}\: |A^2|^\frac{1}{2}\:, \]
and squaring both sides we find that
\[ {\mathcal{L}}_\mu \;\geq\; 0 \spc {\mbox{if }} \quad \mu \leq \frac{1}{2n}\:. \]
If the last inequality is strict, we get existence:
\begin{Thm} \label{thmn2}
If $\mu < \frac{1}{2n}$, the auxiliary variational principle~(\ref{vp})
attains its minimum.
\end{Thm}
If conversely~$\mu > \frac{1}{2n}$, one sees in the example of the matrices
$A_k = k \1$ that~$|A_k^2| = 2 n k^2$, $|A_k|^2 = 4 n^2 k^2$ and thus
${\mathcal{L}}_\mu[A_k] \rightarrow -\infty$ as~$k \rightarrow \infty$. Hence the
action is not bounded below, and we cannot expect the existence of minimizers.

The remaining {\em{critical case}} $\mu=\frac{1}{2n}$ is the most interesting but
also the most difficult case. For notational convenience,
we set~${\mathcal{L}} \equiv {\mathcal{L}}_{1/2n}$
and~${\mathcal{S}} \equiv {\mathcal{S}}_{1/2n}$.
Then our Lagrangian can also be written in the form
\beq \label{L2}
{\mathcal{L}}[A] \;=\; |A^2| - \frac{1}{2n}\: |A|^2 \;=\;
\frac{1}{4n} \sum_{i,j=1}^{2n} \left( |\lambda_i| - |\lambda_j| \right)^2 \:,
\eeq
as is easily verified by multiplying out the last square in~(\ref{L2}).
This shows that the Lagrangian
vanishes only if the~$|\lambda_j|$ are all equal. Thus one can say
qualitatively that the critical variational principle tries to achieve that the
zeros of the characteristic polynomial of~$A$ all have the same absolute
value.

In the critical case we prove the following existence theorem.
\begin{Thm} \label{thmn3}
Suppose that~$(P_k)_{k \in \sN}$ is a minimal sequence of the auxiliary
variational principle~(\ref{vp}) in the critical case~$\mu = \frac{1}{2n}$. Assume that
the local trace is bounded away from zero in the sense that
for suitable~$\delta>0$,
\[ |\Tr(E_x P_k)| \;\geq\; \delta \spc \forall \:k \in \N,\: x \in M\:. \]
Then there exists a minimizer~$P$.
\end{Thm}
Here we need the additional condition that in a minimal sequence
the local trace must not go to zero at any
space-time point. It is an open problem whether this condition is only
a technicality needed in our proof, or whether it is really
necessary for the theorem to hold.

We will prove a general existence theorem (see Theorem~\ref{thm2} below), which is
useful for constructing minimizers under various constraints. As an example,
we here consider homogeneous operators.
\begin{Def} \label{defhomo}
A fermionic projector~$P$ is called {\bf{homogeneous}} if for
any~$x_0, x_1 \in M$ there is a permutation~$\sigma \::\: M \rightarrow M$
with~$\sigma(x_0)=x_1$ and a gauge transformation~$U \in {\mathcal{G}}$ such that
\[ P(\sigma(x), \sigma(y)) \;=\; U\: P(x,y)\: U^{-1}\spc \forall\, x,y \in M\:. \]
\end{Def}
We remark that this definition generalizes the usual notion of ``homogeneity'' as defined
via a symmetry group~$K$ acting transitively on space-time.
Namely, in this case we take for any~$x_0, x_1 \in M$ a group element~$g \in K$
with~$g(x_0)=x_1$ and set~$\sigma(x)=g(x)$, together with
unitary maps~$U_x : E_x(H) \to E_{g(x)}(H)$
which identify the corresponding spinor spaces.
Homogeneous operators seem of physical interest because
the vacuum should be described by a homogeneous fermionic projector.
\begin{Thm} \label{corres}
Consider the auxiliary variational principle~(\ref{vp}) in the critical case~$\mu=\frac{1}{2n}$.
Varying~$P$ in the class of homogeneous fermionic projectors, the action~(\ref{Sdef})
attains its minimum.
\end{Thm}

In the course of our analysis, it will be convenient to consider our variational
principles more generally on operators~$P$ which are not necessarily projectors.
These generalizations are of interest if one considers the
above variational principles on a subspace of~$H$, disregarding the overall
normalization of the fermionic states (for example, one may consider a system
corresponding to a subset of space-time points or modeling only one sector).
In this situation, the restriction of~$P$ to the subspace is no longer a projector.
In order to specify which class of operators~$P$ we want to consider, we need the
following notion.
\begin{Def} \label{defpositive}
A symmetric operator~$A$ on an indefinite inner product space of signature~$(p,q)$
is said to be {\bf{positive}} if
\[ \bra u \:|\: A\, u \ket \;\geq\; 0 \spc \forall \:u \in H\:. \]
\end{Def}
If~$P$ is a projector on a negative definite subspace, then the operator~$(-P)$
is positive because
\beq \label{Pj}
\bra u \:|\: (-P)\, u \ket \;=\;  - \bra u \:|\: P^2 \, u \ket
\;=\; - \bra P u \:|\: P u \ket \;\geq\; 0\:.
\eeq
Therefore, the next definition really extends the class of fermionic projectors
of rank~$f$.
\begin{Def} \label{def28}
An operator~$P$ on an inner product space~$(H, \bra .|. \ket)$ is said to be
{\bf{of class~${\mathcal{P}}^f$}} if
\begin{description}
\item[(i)] The operator~$(-P)$ is positive.
\item[(ii)] The operator~$P$ has trace~$f$ and rank at most~$f$.
\end{description}
\end{Def}

\begin{Thm} \label{thm0}
The variational principle~(\ref{vary}, \ref{constraint}) considered
for~$P \in {\mathcal{P}}^f$ attains its minimum.
\end{Thm}

\begin{Thm} \label{thm1}
For every~$\mu \leq \frac{1}{2n}$, the auxiliary variational principle~(\ref{Sdef})
attains its minimum in~${\mathcal{P}}^f$.
\end{Thm}
We point out that the last theorem also applies in the critical case~$\mu =
\frac{1}{2n}$.

In Def.~\ref{def28}~{\bf{(ii)}}, the operator~$P$ was normalized
by prescribing its trace.
Such a normalization is essential for the auxiliary variational principle in order to
rule out the trivial minimizer~$P=0$. However, for the variational principle~(\ref{vary}),
the constraint~(\ref{constraint}) prevents trivial solutions, and thus it makes sense
to drop the normalization.
\begin{Thm} \label{thmn5}
Consider for any parameters~$\kappa>0$ and~$f \in \N$ the variational
principle~(\ref{vary}, \ref{constraint}), where~$P$ now is a general operator such
that~$(-P)$ is positive and has rank at most~$f$. Then there exists a minimizer~$P$.
It is a stationary point of the action~(\ref{Sdef}, \ref{Ldef}) with the Lagrangian
multiplier chosen such that
\beq \label{range}
{\mathcal{S}}_\mu[P] \;=\; 0\:.
\eeq
\end{Thm}
This theorem is interesting because of the following argument, which explains why
the condition for~$P$ being a projector is fundamental: One may wonder why the
physical fermionic projector was introduced in~\cite{PFP} as a projector.
The only reason was to ensure the correct normalization of the fermionic states,
in accordance with physical observations.
But since physical observations are limited to the low-energy region,
it is a-priori not clear if the physical~$P$ really is a projector, or whether
only its low-energy states give the impression that~$P$ is a projector.
Theorem~\ref{thmn5} gives us information on what would happen if the normalization
condition for~$P$ were dropped. Then the action of the minimizer would vanish~(\ref{range}),
in contradiction to the fact that for a Dirac sea configuration, the
Lagrangian~${\mathcal{L}}_\mu[A_{xy}]$ is strictly positive if the vector~$y-x$ is
timelike (see~\cite[\S5.6]{PFP}).

For the proof of the above theorems we will use the direct method of the
calculus of variations. By starting with different minimal sequences, our
method allows to construct all minimizers.

\section{Discussion and Simple Examples} \label{sec3}
We begin with a few general remarks on the mathematical structure of the
variational principle introduced in the previous section.
First, we point out that the
Lagrangian~${\mathcal{L}}_\mu[A_{xy}]$ is symmetric in its two arguments~$x$
and~$y$, as the following consideration shows. For any two quadratic
matrices~$B$ and~$C$, we choose~$\varepsilon$ not in the spectrum of~$C$ and
set~$C^\varepsilon = C-\varepsilon\1$. Taking the determinant of the
relation $C^\varepsilon (B C^\varepsilon - \lambda) = (C^\varepsilon B - \lambda) C^\varepsilon$,
we can use that the determinant is multiplicative and that~$\det C^\varepsilon \neq 0$
to obtain the equation~$\det(B C^\varepsilon -\lambda) = \det(C^\varepsilon B -\lambda)$. Since
both determinants are continuous in~$\varepsilon$, this equation holds even for
all~$\varepsilon \in \R$, proving the elementary identity
\[ \det(B C -\lambda \1) \;=\; \det(C B -\lambda \1) \:. \]
Applying this identity to the closed chain,
\begin{eqnarray*}
\det(A_{xy} - \lambda \1) &=& \det(P(x,y)\, P(y,x) - \lambda \1) \\
&=& \det(P(y,x)\, P(x,y) - \lambda \1) \;=\; \det(A_{yx} - \lambda \1) \:,
\end{eqnarray*}
we conclude that the operators~$A_{xy}$ and~$A_{yx}$ have the same
characteristic polynomial, and thus
\beq \label{symmetry}
{\mathcal{L}}_\mu[A_{xy}] \;=\; {\mathcal{L}}_\mu[A_{yx}] \spc
\forall \, x,y \in M\:.
\eeq

It is a simple but important observation that a joint unitary transformation
of all projectors,
\beq \label{unit}
E_x \;\to\; U E_x U^{-1} \:, \qquad
P \;\to\; U P U^{-1} \spc {\mbox{with~$U$
unitary}}
\eeq
keeps the action unchanged, because
\begin{eqnarray*}
P(x,y) &\to& U\:P(x,y)\:U^{-1} \:,\spc
A_{xy} \;\to\; U A_{xy} U^{-1} \\
\det (A_{xy} - \lambda \1) &\to&
\det \!\left( U (A_{xy} - \lambda\1)\: U^{-1} \right) \;=\;
\det (A_{xy} - \lambda\1)\:,
\end{eqnarray*}
and so the~$\lambda_j$ stay the same. Such unitary transformations can also
be used to vary the fermionic projector. However, since we want to keep discrete
space-time fixed, we are only allowed to consider unitary
transformations which do not change the space-time projectors,
\beq \label{gauge1}
E_x \;=\; U E_x U^{-1} \spc \forall \:x \in M\:.
\eeq
Then~(\ref{unit}) reduces to the transformation of the fermionic projector
\beq \label{gauge2}
P \;\to\; U P U^{-1}\:.
\eeq
Unitary transformations of the form~(\ref{gauge1}, \ref{gauge2}) are
called {\em{gauge transformations}}.
The conditions~(\ref{gauge1}) mean that~$U$ maps every subspace~$E_x(H)$ into
itself. Hence~$U$ splits into a direct sum of unitary transformations
\beq \label{local}
U(x) \;:=\; U E_x \;:\; E_x(H) \: \rightarrow\: E_x(H) \:,
\eeq
which act ``locally'' on the subspaces associated to the individual space-time
points. Obviously, the gauge transformations form a group, referred to as the
{\em{gauge group}}~${\mathcal{G}}$.
Localizing the gauge transformations according to~(\ref{local}), we obtain
at any space-time point~$x$ the so-called {\em{local gauge group}}.
The local gauge group is the group of
isometries of~$E_x(H)$ and can thus be identified with the group~$U(n,n)$.

One may ask why the space-time projectors are to be kept fixed in our
variational principles. More generally, one could vary both~$P$
and the~$(E_x)_{x \in M}$, fixing only the integer parameters~$f$
and~$n$. Recall that the space-time projectors are equivalently
described by the orthogonal decomposition~(\ref{odecomp}) together
with the condition that the subspaces~$E_x(H)$ should all have
signature~$(n,n)$. For two different sets of space-time projectors, we can find
a unitary transformation which maps the corresponding subspaces~$E_x(H)$
onto each other. Then the transition from one set of space-time projectors
to the other is described by the unitary transformation~$E_x \to
U E_x U^{-1}$. Since such unitary transformations leave the action unchanged
if also the fermionic projector is transformed according to~(\ref{unit}),
it is no loss in generality to fix the space-time projectors throughout.

It is instructive to consider our framework in a concrete basis of~$H$.
Then our inner product can be represented in the form
\[ \bra u \:|\: v \ket \;=\; (u \:|\: S v) \:, \]
where~$(.|.)$ is the canonical scalar product on~$\C^{2mn}$. Here~$S$ is a
Hermitian matrix (meaning that~$(u \:|\: S v) = (S u \:|\: v)\; \forall\:
u,v \in H$), referred to as the {\em{signature matrix}}. By choosing the
basis of~$H$ appropriately, we can arrange that~$S$ is diagonal with
eigenvalues equal to~$\pm 1$. In particular, $S$ is unitary and~$S^2 = \1$.
The signature matrix is useful for calculations. For example,
\[ \bra u \:|\: A v \ket \;=\; (u \:|\: S A v) \;=\;
(A^\dagger S u \:|\: v) \;=\; (S A^\dagger S u \:|\: S v) \;=\;
\bra S A^\dagger S u \:|\: v \ket\:, \]
where the dagger denotes transposition and complex conjugation. Thus
the adjoint can be expressed by
\[ A^* \;=\; S A^\dagger S\:. \]
In particular, a matrix is symmetric if and only if the matrix~$SA$ is Hermitian. As one already sees in the two-dimensional example
\beq \label{ex1}
S \;=\; \left( \! \begin{array}{cc} 1 & 0 \\ 0 & -1 \end{array} \! \right) ,
\spc A \;=\; \left( \! \begin{array}{cc} 1 & 1 \\ -1 & -1 \end{array} \! \right) ,
\eeq
a symmetric matrix in an indefinite inner product space need not be diagonalizable.
This explains why after~(\ref{cc}) we had to speak of ``zeros of the
characteristic polynomial'' and not of ``eigenvalues.'' Note that the
matrix~$A$ in~(\ref{ex1}) is nilpotent and thus~$|A|=0$. This shows that
the spectral weight is not a matrix norm, not even on symmetric operators.
We remark that it seems impossible to introduce any other basis independent
matrix norm; in particular, the analogue of the Hilbert-Schmidt norm
$(\Tr(A^* A))^{\frac{1}{2}}$ vanishes in the example~(\ref{ex1}).
Even if a symmetric matrix {\em{is}} diagonalizable, its eigenvalues are in general not
real, as can be seen in the example
\beq \label{ex2}
S \;=\; \left( \! \begin{array}{cc} 1 & 0 \\ 0 & -1 \end{array} \! \right) ,
\spc A \;=\; \left( \! \begin{array}{cc} 0 & 1 \\ -1 & 0 \end{array} \! \right) .
\eeq
At least, the calculation
\[ \overline{ \det(A-\lambda\1) } \;=\;
\det \!\left( A^\dagger - \overline{\lambda} \right) \;=\;
\det \!\left( S(A^\dagger - \overline{\lambda})S \right) \;=\;
\det \!\left( A^* - \overline{\lambda} \right) \;=\;
\det \!\left( A - \overline{\lambda}\1 \right) \]
shows that the characteristic polynomial of a symmetric matrix~$A$ has real
coefficients. In other words, the non-real~$\lambda_j$ always appear in
complex conjugate pairs.

Using the above matrix representations, we can now consider a few simple examples.
We restrict attention to the auxiliary variational principle~(\ref{vp}) in the
critical case~$\mu=\frac{1}{2n}$ because this case seems most interesting.
The examples are generalized in a straightforward way to the other cases
and to the variational principle~(\ref{vary}).
We begin with the case~$m=1$ of
one space-time point. In this case, the only space-time projector~$E$
is the identity, and the sum over the space-time points in~(\ref{Sdef})
drops out. Thus
\[ {\mathcal{S}} \;=\; |A^2| - \frac{1}{2n}\: |A|^2 \]
with~$A=P^2$. Using that~$P$ is idempotent and that its only
non-vanishing eigenvalue is one with multiplicity~$f$, we find that
\[ {\mathcal{S}} \;=\; |P| - \frac{1}{2n}\: |P|^2 \;=\; f - \frac{f^2}{2n}\:. \]
Hence the action is unchanged if the fermionic projector is varied.
This can also be understood from the fact that with only one space-time point,
the condition~(\ref{gauge1}) is trivial, and therefore any variation
of~$P$ can be realized as a gauge transformation~(\ref{gauge2}).
The situation becomes more interesting with two space-time points,
as the next example shows.

\begin{Example} \label{example1} \em
Choose~$M=\{ 1,2 \}$ with spin dimension~$(1,1)$ and $f=1$.
Then~$H$ is 4-dimensional, and by choosing a suitable basis we can arrange that
\beq \label{exdst}
S \;=\; \left( \begin{array}{cccc} 1 & 0 & 0 & 0 \\
0 & -1 & 0 & 0 \\ 0 & 0 & 1 & 0 \\ 0 & 0 & 0 & -1
\end{array} \right) ,\spc
E_1 \;=\; \left( \begin{array}{cc} \1 & 0 \\
0 & 0
\end{array} \right) ,\quad
E_2 \;=\; \left( \begin{array}{cc} 0 & 0 \\ 0 & \1
\end{array} \right) ,
\eeq
where for~$E_{1\!/\!2}$ we used a block matrix notation (thus every
matrix entry stands for a $2 \times 2$-matrix). Again in this block matrix
notation, the gauge transformations~(\ref{gauge1}) are of the form
\beq \label{exgauge}
U \;=\; \left( \begin{array}{cc} U_1 & 0 \\ 0 & U_2
\end{array} \right) \:,
\eeq
where~$(U_x)_{x \in M}$ are two independent ``local'' unitary transformations
on~$E_x(H)$ of the form
\[ U_x \;=\; e^{i \alpha} \left( \begin{array}{cc} e^{i \beta}\, \cosh \vartheta
& e^{i \gamma}\, \sinh \vartheta \\ e^{-i \gamma}\, \sinh \vartheta &
e^{-i \beta}\, \cosh \vartheta
\end{array} \right) \qquad {\mbox{with $\alpha, \beta, \gamma, \vartheta \in \R$}}. \]
Thus the local gauge group is~$U(1,1)$, and
the gauge transformations~(\ref{exgauge}) are elements
of the gauge group~${\mathcal{G}} = U(1,1) \otimes U(1,1)$.

Since we consider a system of one particle ($f=1$), the fermionic projector~$P$
must be a projector on a one-dimensional, negative definite subspace. It is
convenient to write~$P$ using bra/ket-notation as
\beq \label{Prep}
P \;=\; -|\,u \ket \bra u\,| \spc {\mbox{with}} \spc
\bra u \:|\: u \ket \;=\; -1\:.
\eeq
A possible choice is
\beq \label{exP1}
u \;=\; \left( \!\begin{array}{c}
0 \\ 1 \\ 0 \\ 0 \end{array}\! \right) \spc {\mbox{and thus}} \spc
P \;=\; \left( \begin{array}{cccc}
0 & 0 & 0 & 0 \\ 0 & 1 & 0 & 0 \\0 & 0 & 0 & 0 \\0 & 0 & 0 & 0
\end{array} \right).
\eeq
A short calculation yields that~$|A_{11}|=|A_{11}^2|=1$,
and all other~$A_{ij}$ vanish. Thus
\[ {\mathcal{S}} \;=\; {\mathcal{L}}(A_{11})
\;=\; |A_{11}^2| - \frac{1}{2}\:|A_{11}|^2 \;=\; \frac{1}{2}\:. \]

It turns out that the above~$P$ is not a minimizer. Namely, choosing
\beq \label{exP2}
u \;=\; \frac{1}{\sqrt{2}} \left(\! \begin{array}{c} 0 \\ 1 \\ 0 \\ 1 \end{array}
\!\right) , \qquad
P \;=\; \frac{1}{2} \left( \begin{array}{cccc}
0 & 0 & 0 & 0 \\ 0 & 1 & 0 & 1 \\0 & 0 & 0 & 0 \\0 & 1 & 0 & 1
\end{array} \right) ,
\eeq
we get a smaller value for the action,
\begin{eqnarray*}
|A_{ij}|^2 &=&  |A_{ij}^2| \;=\; \frac{1}{16} \quad {\mbox{for all~$i,j \in M$}} \\
{\mathcal{S}} &=& 4\: {\mathcal{L}}(A_{11})
\;=\; 4 \left(\frac{1}{16} - \frac{1}{2 \cdot 16} \right) \;=\;
\frac{1}{8}\:.
\end{eqnarray*}

Let us verify that this is indeed the minimum. We represent a
general~$P$ in the form~(\ref{Prep}).
Since at least one of the inner products~$\bra u \,|\, E_1\, u \ket$
or~$\bra u \,|\, E_2 \,u \ket$ must be negative, we must distinguish
the two cases where these two inner products either have the opposite sign
or are both non-positive.
In the first case, we can assume that~$\bra u | E_1\, u \ket >0$
and~$\bra u | E_2 \,u \ket <0$. Using the
gauge freedom, we can arrange that~$u$ is of the
form~$u=(\sinh \varphi, 0, 0, \cosh \varphi)$ with~$\varphi \in \R$.
A short calculation yields that
\begin{eqnarray*}
|A_{11}|^2 &=& |A_{11}^2| \;=\; \sinh^8 \varphi \:,\qquad
|A_{22}|^2 \;=\; |A_{22}^2| \;=\; \cosh^8 \varphi \\
|A_{12}|^2 &=& |A_{12}^2| \;=\; |A_{21}|^2 \;=\; |A_{21}^2|
\;=\; \sinh^4 \varphi\: \cosh^4 \varphi \\
{\mathcal{S}} &=& \sum_{i,j \in M} |A_{ij}^2| - \frac{1}{2}\:|A_{ij}|^2 \;=\;
\frac{1}{2} \left(\cosh^4 \varphi + \sinh^4 \varphi \right)^2
\;\geq\; \frac{1}{2} \;>\; \frac{1}{8}\:.
\end{eqnarray*}

In the remaining case when the inner products~$\bra u \,|\, E_1\, u \ket$
and~$\bra u \,|\, E_2 \,u \ket$ are both non-positive,
we can use the gauge freedom~(\ref{exgauge}) to arrange that~$u$ is
of the form~$u=(0, \cos \varphi, 0, \sin \varphi)$ with~$\varphi \in
[0, 2 \pi)$. It follows that
\begin{eqnarray*}
|A_{11}|^2 &=& |A_{11}^2| \;=\; \cos^8 \varphi \:,\qquad
|A_{22}|^2 \;=\; |A_{22}^2| \;=\; \sin^8 \varphi \\
|A_{12}|^2 &=& |A_{12}^2| \;=\; |A_{21}|^2 \;=\; |A_{21}^2|
\;=\; \sin^4 \varphi\: \cos^4 \varphi \\
{\mathcal{S}} &=& \sum_{i,j \in M} |A_{ij}^2| - \frac{1}{2}\:|A_{ij}|^2 \;=\;
\frac{1}{2} \left(\cos^4 \varphi + \sin^4 \varphi \right)^2 \;=\;
\frac{1}{2} \left(2 \sin^4 \varphi - 2 \sin^2 \varphi + 1 \right)^2 \:,
\end{eqnarray*}
and the last function really attains its minimum when~$\sin^2 \varphi = 1/2$.
\QEDrem
\end{Example}

In the above example, our variational principle has, up to gauge transformations,
a unique minimum. The fact that the configuration~(\ref{exP1}), where the particle
is localized at the first space-time point, is not optimal can be understood
qualitatively by saying that our variational principle ``tends to spread out
particles in space-time.'' We will quantify this observation later
(see Lemma~\ref{lemmamon}); it will be important in our analysis.
We also point out that the local gauge group is {\em{non-compact}}, and that the set of
gauge-equivalent minima~$U P U^{-1}$ with~$P$ and~$U$ according to~(\ref{exP2}, \ref{exgauge})
form an {\em{unbounded}} family of matrices. This explains why minimizers cannot
be constructed with simple compactness arguments.

We next consider a system of two particles.
\begin{Example} \label{example2} \em
Choose~$M=\{ 1,2 \}$ with spin dimension~$(1,1)$ and $f=2$. Thus
the discrete space-time is the same as in Example~\ref{example1};
it is again described by~(\ref{exdst}). However, $P$ now maps onto a
two-dimensional negative subspace of~$H$. An example for~$P$ is obtained
by localizing one particle at each space-time point,
\beq \label{Plocal}
P \;=\; \left( \begin{array}{cccc}
0 & 0 & 0 & 0 \\ 0 & 1 & 0 & 0 \\0 & 0 & 0 & 0 \\0 & 0 & 0 & 1
\end{array} \right) .
\eeq
This is indeed the minimizer, as the following
consideration shows. Let~$P$ be a general femionic projector of rank two.
We first consider the case that both operators~$E_1 P E_1$
and~$E_2 P E_2$ have a non-trivial kernel. We set~$\alpha_x = \Tr(E_x P)$.
Then~$\alpha_1+\alpha_2=2$ and~$\Tr((E_x P)^2) = \Tr(E_x P)^2 =
\alpha_x^2$. Furthermore,
\[ \alpha_1 \;=\; \Tr (E_1 P^2) \;=\;
\Tr (E_1 \,P \,E_1 \,P) + \Tr (E_1 \,P \,E_2 \,P)
\;=\; \alpha_x^2 + \Tr (E_1 \,P \,E_2 \,P) \]
and similarly for~$\alpha_2$. It follows that
\[ \alpha_1 - \alpha_2 \;=\; \alpha_1^2 - \alpha_2^2 \;=\;
(\alpha_1-\alpha_2)(\alpha_1 + \alpha_2) \;=\; 2\, (\alpha_1 - \alpha_2)\:. \]
The only solution to the above equations
is~$\alpha_1=\alpha_2=1$. Using that~$P$ is a projector
on a two-dimensional negative definite subspace, it is easily
verified that~$P$ is gauge equivalent to the fermionic
projector~(\ref{Plocal}).

It remains to consider the case that for example~$E_1 P E_1$
has a trivial kernel. Then its characteristic polynomial has two
non-vanishing roots. Anticipating results of Section~\ref{sec4},
the operator~$(-E_1 P E_1)$ is positive (Lemma~\ref{lemmares}~{\bf{(i)}}),
and the two non-zero roots have opposite signs (Lemma~\ref{lemmaspec}).
Thus~$E_1 P E_1$ can be diagonalized. More precisely, we can choose
a new pseudo-orthonormal basis in~$E_1(H)$ such that the operator~$P$
takes the form
\[ P \;=\; \left( \begin{array}{cccc}
a & 0 & * & * \\ 0 & b & * & * \\ \ast & * & * & * \\ \ast & * & * & *
\end{array} \right)
\quad {\mbox{with~$a<0$ {\mbox{and}} $b>0$}}, \]
where the stars denote arbitrary entries.
The first column of this matrix is a vector in the negative definite subspace~$P(H)$.
Using a gauge transformation in~$E_2(H)$, we can arrange that this vector
is of the form~$u = (a,0,0,c)$ with~$|c|>|a|$. Using that~$Pu=u$ and~$P^*=P$,
we can arrange by a suitable gauge transformation in~$E_2(H)$ that~$P$ is of the form
\[ P \;=\; \left( \begin{array}{cccc}
-\sinh^2 \alpha & 0 & 0 & \sinh \alpha \cosh \alpha \\ 0 & b & * & 0 \\
0 & * & * & 0 \\ -\sinh \alpha \cosh \alpha & 0 & 0 & \cosh^2 \alpha
\end{array} \right) \quad {\mbox{with~$\alpha \neq 0$}}.\]
Again using that~$P$ is a projector on a negative definite
subspace, we can arrange that
\[ P \;=\; \left( \begin{array}{cccc}
-\sinh^2 \alpha & 0 & 0 & \sinh \alpha \cosh \alpha \\
0 & \cosh^2 \beta & -\sinh \beta \cosh \beta & 0 \\
0 & \sinh \beta \cosh \beta & -\sinh^2 \beta & 0 \\
-\sinh \alpha \cosh \alpha & 0 & 0 & \cosh^2 \alpha
\end{array} \right) . \]
In the limiting case~$\alpha=\beta=0$, this formula also
includes~(\ref{Plocal}).
The Lagrangian corresponding to this fermionic projector
is computed to be
\begin{eqnarray*}
{\mathcal{L}}[A_{11}] &=& \frac{1}{2} \left(\sinh^4 \alpha - \cosh^4 \beta \right)^2 \\
{\mathcal{L}}[A_{22}] &=& \frac{1}{2} \left(\sinh^4 \beta - \cosh^4 \alpha \right)^2 \\
{\mathcal{L}}[A_{12}] \;=\; {\mathcal{L}}[A_{21}]
&=& \frac{1}{2} \left(\cosh^2 \alpha \sinh^2 \alpha - \cosh^2 \beta \sinh^2 \beta \right)^2 .
\end{eqnarray*}
Adding these terms and using trigonometric identities for the hyperbolic functions,
we obtain for the action the expression
\[ {\mathcal{S}}[P] \;=\; \frac{1}{8} \Big( 2 + (\cosh 2\alpha - \cosh 2\beta)^2 \Big)
\left( \cosh 2\alpha + \cosh 2\beta \right)^2 \:, \]
and this function clearly attains its unique
absolute minimum at~$\alpha = 0 = \beta$.
\QEDrem
\end{Example}

The most interesting case is when the number of particles is large,
but still much smaller than the number of space-time points.
\begin{Example} \label{example3} \em
Choose~$1 \ll f \ll m$ with spin dimension~$(1,1)$. In this case, we
can represent discrete space-time by the following matrices,
\[ S \;=\; \left( \begin{array}{ccccc}
1 & 0 &&& \\ 0 & -1 &&& \\ && 1 & 0 & \\ && 0 & -1 & \\
&& && \ddots
\end{array} \right) \]
and
\[ E_1 = \left( \begin{array}{ccc}
\1 & & \\ &0 & \\ && \ddots \end{array} \right) , \quad
E_2 = \left( \begin{array}{ccc}
0 & & \\ & \1 & \\ && \ddots \end{array} \right) ,\ldots, \;E_m\:.  \]
One possibility to choose~$P$ is to localize each of the~$f$
particles similar to~(\ref{Plocal}) at one of the space-time points.
However, the resulting value for the action
\[ {\mathcal{S}} \;=\; \frac{f}{2} \]
is certainly not minimal. It is better if, in analogy to~(\ref{exP2}),
each particle is evenly spread over~$m/f$ space-time points (we here assume
for simplicity that~$m/f$ is an integer). A short calculation yields
\[ {\mathcal{S}} \;=\; \frac{f}{2} \left( \frac{f}{m} \right)^2 \:. \]
There is no reason why this configuration should be optimal.
It is completely unknown how the minimizer looks like in general.
\QEDrem
\end{Example}

The case of physical interest is spin dimension~$(2,2)$
(or more generally~$(2N, 2N)$ with $N \geq 1$), because in this
case the vectors of~$H$ can be identified with the Dirac wave functions
of relativistic quantum mechanics.
We expect the general structure of the minima to be very complicated.
Our qualitative picture is that
the minimizers should induce relations between
the discrete space-time points which for large~$m$ and~$f$ should
correspond to specific geometric configurations of the space-time points.
As explained in~\cite[{\S}5.6]{PFP}, such relations should, in a suitable limit
in which discrete space-time goes over to a continuum space-time,
give the causal structure of a Lorentzian manifold.

\section{Positive Operators, Lower Bounds for the Lagrangian} \label{sec4}
We return to the concept of a positive operator as introduced
in Def.~\ref{defpositive}.
Expressed with the signature matrix, we can say that a self-adjoint operator~$A$
on an inner product space~$H$ of signature~$(p,q)$ is positive
if and only if the matrix~$SA$ is positive semi-definite on~$\C^{p+q}$
endowed with the standard Euclidean scalar product.
To avoid confusion, we point out that the statements ``$A$ is positive'' and
``the image of~$A$ is positive'' are completely different. In the
two-dimensional examples
\beq \label{ex3}
S \;=\; \left( \! \begin{array}{cc} 1 & 0 \\ 0 & -1 \end{array} \! \right) ,
\spc A_1 \;=\; \left( \! \begin{array}{cc} 1 & 0 \\ 0 & -1 \end{array} \! \right) , \quad
A_2 \;=\; \left( \! \begin{array}{cc} 0 & 0 \\ 0 & -1 \end{array} \! \right) ,
\eeq
the operator~$A_1$ is positive, although its image has signature~$(1,1)$. The operator~$A_2$ is
also positive, but its image is negative. The last example also shows that
the trace of a positive operator can be negative.
At least, the argument~(\ref{Pj}) shows that a {\em{projector}} on a positive
subspace is a positive operator.

We now collect a few elementary but useful properties of positive operators.
\begin{Lemma} \label{lemmares}
Suppose that~$A$ is a positive operator on~$H$. Then
\begin{description}
\item[(i)] If~$Q$ is a projector in~$H$, the operator
$QAQ$ is again positive.
\item[(ii)] For all~$u, v \in H$,
\beq \label{schwarz}
\left| \bra u \:|\: A\, v \ket \right| \;\leq\;
\sqrt{\bra u \:|\: A\, u \ket} \;\sqrt{\bra v \:|\: A\, v \ket} \:.
\eeq
\end{description}
\end{Lemma}
{\Proof} Part~{\bf{(i)}} is obvious because~$\bra u \,|\, QAQ\, u \ket =
\bra Q u \,|\, A \, Q u \ket \geq 0$. Part~{\bf{(ii)}} can be regarded as the
Schwarz inequality for the positive semi-definite inner product~$(.|.)_A := \bra .|A. \ket$.
The proof is almost as simple as in scalar product spaces: First note that for
all~$a,b \in H$,
\[ 0 \;\leq\; (a-b \:|\: a-b )_A \;=\;
( a \:|\: a)_A \:+\: (b \:|\: b)_A \:-\:
2 \,{\mbox{Re}} \,(a \:|\: b )_A \:. \]
By changing the phase of the vector~$a$, we can arrange that~$( a \,|\, b )_A \geq 0$. Thus
\beq \label{sch1}
2 \left|(a \:|\: b )_A \right| \;\leq\;
( a \:|\: a)_A \:+\: (b \:|\: b )_A \spc \forall \:a,b \in H\: .
\eeq

Suppose that~$(u \,|\, u)_A=0$. Then applying~(\ref{sch1}) for~$a=u/\varepsilon$
and~$b=\epsilon u$ with a parameter~$\epsilon>0$ gives
\[ 2 \left|(u \:|\: v)_A \right| \;\leq\; \varepsilon^2\: (v \:|\: v)_A \:, \]
and letting~$\varepsilon \to 0$, we see that~(\ref{schwarz}) is
trivially satisfied. The same argument applies if~$(v \,|\, v)_A=0$.
In the remaining case~$(u \,|\, u)_A \neq 0$ and~$(v \,|\, v)_A \neq 0$, we apply~(\ref{sch1})
with
\[ a \;=\; \left( \frac{(v \:|\: v)_A}{(u \:|\: u)_A} \right)^\frac{1}{4} u \:, \spc
b \;=\; \left( \frac{(u \:|\: u)_A}{(v \:|\: v)_A} \right)^\frac{1}{4} v \:. \]

\vspace*{-.7cm}
\QED

Compared to the situation for general
symmetric operators as explained after~(\ref{ex1}), positive
operators have nice spectral properties, as the
following approximation argument shows.
\begin{Lemma} \label{lemmaspec}
A positive operator~$A$ on an indefinite inner product space of
signature~$(p,q)$ has a purely real spectrum. The
zeros~$(\lambda_j)_{j=1,\ldots,p+q}$ of its
characteristic polynomial (again counted with multiplicities) can be ordered
as follows,
\[ \lambda_1 \leq \cdots \leq \lambda_q \;\leq\; 0
\;\leq\; \lambda_{q+1} \leq \cdots \leq \lambda_{p+q}\:. \]
\end{Lemma}
{\Proof}
We choose a matrix representation with signature matrix~$S$
and set~$A^\varepsilon = A + \varepsilon S$.
Clearly, the matrices~$A^\varepsilon$ converge to~$A$ as
$\varepsilon \to 0$. Since the spectrum is continuous
in~$\varepsilon$, it suffices to prove the lemma for
the matrix~$A^\varepsilon$ and any~$\varepsilon>0$.

The matrix~$A^\varepsilon$ is symmetric and strictly positive in the sense that
for all $u \neq 0$,
\[ \bra u \:|\: A^\varepsilon u \ket \;=\;
\bra u \:|\: A\, u \ket \:+\: \varepsilon\, \bra u \:|\: S\, u \ket \;\geq\;
\varepsilon \,\bra u \:|\: S\, u \ket \;=\; 
\varepsilon \,(u \:|\: u) \;>\; 0\:. \]
Hence we can introduce a scalar product by
\[ (u \:|\: v)_{A^\varepsilon} \;:=\; \bra u \:|\: A^\varepsilon v \ket \:. \]
Since the operator~$A^\varepsilon$ is symmetric and commutes with itself,
it is clearly self-adjoint in the Hilbert space~$(H, (.|.)_{A^\varepsilon})$. Thus
we can choose an eigenvector basis~$(u_j)_{j=1,\ldots,p+q}$.
The corresponding eigenvalues~$\lambda_j$ satisfy the identity
\[ \lambda_j\: \bra u_j \:|\: u_j \ket \;=\;
\bra u_j \:|\: A^\varepsilon u_j \ket \;=\; (u_j \:|\: u_j)_{A^\varepsilon} \;>\; 0\:. \]
Thus~$p$ of the eigenvalues are positive, whereas the
other~$q$ eigenvalues are negative.  \QED
Note that a positive operator is in general not diagonalizable,
as the example~(\ref{ex1}) shows.

The above lemmas can be used to get lower estimates of our Lagrangian~(\ref{Ldef})
and the corresponding action,
which shed some light on the mathematical behavior of our variational principle.
We consider the case~$\mu \leq \frac{1}{2n}$. Using the elementary inequality
\beq \label{dec}
{\mathcal{L}}_\mu[A] \;=\; {\mathcal{L}}[A] + \left(\frac{1}{2n} - \mu \right)
|A|^2 \;\geq\; {\mathcal{L}}[A] \spc {\mbox{if }} \mu \leq \frac{1}{2n}\:,
\eeq
we may restrict attention to the critical case~$\mu=\frac{1}{2n}$.
It is obvious from~(\ref{loctr}) that the local trace is non-zero at least at
some~$x \in M$.
The next lemma shows that the Lagrangian of~$A_{xx}$ can be bounded below by expressions
involving the local trace at~$x$.
\begin{Prp} \label{prplower}
Let~$P$ be a symmetric operator on~$(H, \bra .|. \ket)$ such that~$(-P)$
is positive. Then, using again the notation~(\ref{cc}, \ref{Ldef}),
\begin{eqnarray}
{\mathcal{L}}[A_{xx}] &\geq& \frac{|\Tr(E_x P)|^4}{256\, n^3} \label{Llb1} \\
{\mathcal{L}}[A_{xx}] &\geq& \frac{1}{4n}\:|\Tr(E_x P)|^2
\;\inf \sigma \!\left(A_{xx} \big|_{E_x(H)} \right) .
\label{Llb2}
\end{eqnarray}
\end{Prp}
{\Proof} According to Lemma~\ref{lemmares}~{\bf{(i)}}, the operator~$(-P(x,x)) \::\:
E_x(H) \rightarrow E_x(H)$ is positive. Lemma~\ref{lemmaspec} tells us that the
zeros of the characteristic polynomial of~$P(x,x)$, which we denote
by~$(\nu_j)_{j=1,\ldots,2n}$, are all real and have the ordering
\beq \label{morder}
\nu_1 \leq \cdots \leq \nu_n \;\leq\; 0 \;\leq\;
\nu_{n+1} \leq \cdots \leq \nu_{2n} \:.
\eeq
This allows us to write the local trace as follows,
\beq \label{tricky}
\Tr (E_x P) \;=\; \sum_{j=1}^{2n} \nu_j \;=\;
\sum_{i=n+1}^{2n} |\nu_i| - \sum_{j=1}^n |\nu_j|
\;=\; \frac{1}{n} \sum_{i=n+1}^{2n} \sum_{j=1}^n
\left( |\nu_i| - |\nu_j| \right) ,
\eeq
where the last equality is obvious if one notices that when for example adding up
the~$|\nu_i|$, the sum over~$j$ can be carried out giving
a factor~$n$. We now take absolute values and increase the right side
by taking more summands,
\beq \label{ltb}
|\Tr (E_x P)| \;\leq\; \frac{1}{n} \sum_{i,j=1}^{2n}
\Big| |\nu_i| - |\nu_j| \Big| .
\eeq
Now we can proceed with H{\"o}lder's inequality to obtain
\begin{eqnarray}
|\Tr (E_x P)| &\leq& \frac{1}{n} \left( 4n^2 \right)^\frac{1}{2}
\left( \sum_{i,j=1}^{2n} \Big| |\nu_i| - |\nu_j| \Big|^2 \right)^\frac{1}{2}
=\; 2 \left( \sum_{i,j=1}^{2n} \Big| |\nu_i| - |\nu_j| \Big|^2 \right)^\frac{1}{2} \label{es1} \\
|\Tr (E_x P)| &\leq& \frac{1}{n} \left( 4n^2 \right)^\frac{3}{4}
\left( \sum_{i,j=1}^{2n} \Big| |\nu_i| - |\nu_j| \Big|^4 \right)^\frac{1}{4}
=\; \sqrt{8 n} \left( \sum_{i,j=1}^{2n} \Big| |\nu_i| - |\nu_j| \Big|^4 \right)^\frac{1}{4}
\!\!\!\! .
\label{es2}
\end{eqnarray}

Since the matrix~$A_{xx}$ is the square of~$P(x,x)$, the zeros of its
characteristic polynomial, again denoted by~$(\lambda_j)_{j=1,\ldots,2n}$,
satisfy the relations
\beq \label{lammu}
0 \;\leq\; \lambda_j \;=\; |\nu_j|^2\spc \forall \:j=1,\ldots,2n\:.
\eeq
Using the formula~(\ref{L2}) for the critical Lagrangian, it follows that
\beq \label{L3}
4n\, {\mathcal{L}}(A_{xx}) \;=\; \sum_{i,j=1}^{2n} (\lambda_i-\lambda_j)^2
\;=\; \sum_{i,j=1}^{2n} (|\nu_i|-|\nu_j|)^2\:(|\nu_i|+|\nu_j|)^2\:.
\eeq
The last expression can be bounded from below in two ways. Either we use the
inequality
\[ (|\nu_i|+|\nu_j|)^2 \;\geq\; (|\nu_i|-|\nu_j|)^2 \]
and apply~(\ref{es2}) to obtain~(\ref{Llb1}). Or we use the estimate
\[ \sum_{i,j=1}^{2n} (|\nu_i|-|\nu_j|)^2\:(|\nu_i|+|\nu_j|)^2 \;\geq\;
4\: \min_j |\nu_j|^2 \sum_{i,j=1}^{2n} (|\nu_i|-|\nu_j|)^2 \]
together with~(\ref{lammu}) and~(\ref{es1}), giving~(\ref{Llb2}).
\QED

The inequality~(\ref{Llb1}) immediately gives a positive lower bound for the action.
\begin{Corollary} \label{cor1}
For every $P \in {\mathcal{P}}^f$,
the critical action satisfies the inequality
\[ {\mathcal{S}}[P] \;\geq\; \frac{f^4}{256\, n^3\, m^3}\:. \]
\end{Corollary}
{\Proof} We first apply H{\"o}lder's inequality in~(\ref{loctr}),
\beq \label{fb}
f \;\leq\; m^\frac{3}{4} \left( \sum_{x \in M} |\Tr(E_x P)|^4 \right)^\frac{1}{4}\:.
\eeq
Dropping the contributions for~$x \neq y$ in~(\ref{Sdef}), we obtain
the lower bound
\beq \label{drop}
{\mathcal{S}}[P] \;=\; \sum_{x,y \in M} {\mathcal{L}}[A_{xy}]
\;\geq\; \sum_{x \in M} {\mathcal{L}}[A_{xx}] \:,
\eeq
and using~(\ref{Llb1}) and~(\ref{fb}) gives the claim.
\QED

We point out that for the estimates of Proposition~\ref{prplower}
it is crucial that the maximal dimensions of the positive and negative
definite subspaces of~$E_x(H)$ coincide. If we considered more general
discrete space-times with spin dimension~$(p,q)$, then in the case~$p \neq q$ the last transformation
in~(\ref{tricky}) would no longer be valid, and the statements of
Lemma~\ref{prplower} and Corollary~\ref{cor1} would break down.
This can be seen most easily in the extreme example of spin dimension~$(0,q)$,
where by ``localizing'' $q$ particles similar to~(\ref{exP1}) at the
space-time point~$x$ we could arrange that~$P(x,x)=\1_{|E_x(H)}$.
Then~$A_{xx}$ would be the identity, and~${\mathcal{L}}[A_{xx}]$ would vanish,
although the local trace~$\Tr(E_xP)$ would be equal to~$q$.
By localizing all particles in this way at individual space-time points,
we could construct minimizers of the action which are not particularly
interesting.
This consideration is the reason why in this paper we only consider systems of
spin dimension~$(n,n)$. We feel that, apart from their physical significance,
these systems are the ones for which the minimizers of our variational principle
should have the most interesting mathematical structure.

\section{A Lower Bound for the Local Trace} \label{sec6}
In this section we shall analyze how the infimum of our action depends on the
number of space-time points. This will lead us to an estimate for the local
trace of~$P$ (Proposition~\ref{prpmin}), which is needed for the proof of
Theorem~\ref{thm1} in the critical case (the reader not interested in
Theorem~\ref{thm1} may skip this section).

For fixed spin dimension~$(n,n)$ and a fixed
number of particles~$f$, we consider for any~$m \in \N$ a discrete
space-time~$(H, \bra .|. \ket, (E_x)_{x \in M})$ with~$m=\#M$ (note that this discrete
space-time is unique up to isomorphisms). We define for any
fixed~$\mu \leq \frac{1}{2n}$ the quantities
\beq \label{IJdef}
\left. \begin{array}{rcl}
I(f,m) &=& \inf \{ {\mathcal{S}}_\mu[P] \:|\: P \in {\mathcal{P}}^f \} \\
J(f,m) &=& \inf \{ {\mathcal{S}}_\mu[P] \:|\: {\mbox{$P$ fermionic projector}} \}
\end{array} \right\} \:.
\eeq
In the case~$f > mn$, when the set of fermionic projectors is empty, we
set~$J(f,m)=\infty$.
The functions~$I$ and~$J$ are strictly positive by Corollary~\ref{cor1}.
Also, it is obvious that~$I(f,m) \leq J(f,m)$. Apart from simple examples as considered
in Section~\ref{sec3}, nothing is known about the values of~$I(f,m)$ and~$J(f,m)$.
In particular, it would be interesting to know whether~$I(f,m)$ is always
strictly smaller than~$J(f,m)$.

Our next lemma shows that the functions~$I(f,m)$ and~$J(f,m)$ are strictly decreasing
in the parameter~$m$. This can be understood from the fact that if~$m$ is increased,
the particles can spread out over more space-time points, making the
infimum of the action smaller.
\begin{Lemma} \label{lemmamon}
The functions~$I$ and~$J$ defined by~(\ref{IJdef}) satisfy the inequalities
\beq \label{IJin}
I(f,m+1) \;\leq\; \left(1 - \frac{3}{4m} \right) I(f,m) \:,\spc
J(f,m+1) \;\leq\; \left(1 - \frac{3}{4m} \right) J(f,m) \:.
\eeq
\end{Lemma}
{\Proof} Let~$P$ be an operator of class~${\mathcal{P}}^f$ in a discrete
space-time~$(H, \bra .|. \ket, (E_x)_{x \in M})$ with~$M=\{ 1,\ldots, m \}$.
Introducing a discrete space-time~$(\hat{H}, \bra .|. \ket, \hat{M})$
where~$\hat{M} = \{0,\ldots,m \}$ consists of one more space-time point,
there is a unitary transformation~$U$ from~$H$ to the subspace
$K = \oplus_{x=1}^m \hat{E}_x(\hat{H})$ of~$\hat{H}$ which maps the space-time
projectors~$E_x$ to the~$\hat{E}_x$ in the sense that
$E_x = U^{-1} \hat{E}_x U$ for all~$x=1,\ldots,m$. In other words, we
can identify~$(H, \bra .|. \ket, (E_x)_{x \in M})$ with the discrete
space-time~$(K, \bra .|. \ket, (\hat{E}_x)_{x \in M})$. Using this identification,
the operator~$P$ maps~$K$ to itself, and extending it by zero to~$\hat{E}_0(\hat{H})$,
we obtain an operator
\[ P \::\: \hat{H} \rightarrow \hat{H} \spc {\mbox{with}} \spc
E_0 \,P \;=\; 0 \;=\; P\, E_0 \:. \]
Since~$P(x,y)$ vanishes when~$x=0$ or~$y=0$, the action of~$P$ is given by
\[ {\mathcal{S}}_\mu[P] \;=\; \sum_{x, y \in M} {\mathcal{L}}_\mu[A_{xy}] \:, \]
and this also shows that our reinterpretation of~$P$ did not change its action.

Our method is to construct a unitary transformation~$V\,:\, \hat{H} \rightarrow \hat{H}$
such that the action of the operator
\beq \label{hPdef}
\hat{P} \;:=\; V\:P\: V^{-1}
\eeq
is strictly smaller than that of~$P$. First, in
\[ {\mathcal{S}}_\mu[P] \;=\;
\sum_{x \in M} \left( \sum_{y \in M} {\mathcal{L}}_\mu[A_{xy}] \right) \]
we choose a point~$x \in M$ for which the inner sum is maximal. Then
\beq \label{Lbound}
\sum_{y \in M} {\mathcal{L}}_\mu[A_{xy}] \;\geq\; \frac{{\mathcal{S}}_\mu[P]}{m}\:.
\eeq
We choose~$V$ such that it is the identity on the subspaces~$\hat{E}_y(\hat{H})$
for~$y \not \in \{0, x\}$, whereas on the subspace
$\hat{E}_0(\hat{H}) \oplus \hat{E}_x(\hat{H})$ it has in block matrix notation
the form
\[ V \;=\; \frac{1}{\sqrt{2}} \left( \!\begin{array}{cc} \1 & \1 \\ -\1 & \1 \end{array}\!
\right) \:, \spc
V^{-1} \;=\; \frac{1}{\sqrt{2}} \left( \!\begin{array}{cc} \1 & -\1 \\ \1 & \1 \end{array}\!
\right) . \]
A short calculation shows that the discrete kernels of~$P$ and~$\hat{P}$ are related
by
\[ \left\{ \begin{array}{rcll}
\hat{P}(y,z) &=& P(y,z) & {\mbox{if $y,z \not \in \{0, x\}$}} \\[.3em]
\hat{P}(y,z) &=& \displaystyle \frac{1}{\sqrt{2}}\: P(x,z) \:,\;
\hat{P}(z,y) \;=\; \frac{1}{\sqrt{2}}\: P(z,x) &
{\mbox{if~$y \in \{0,x\}$ and~$z \not \in \{0, x\}$}} \\[1em]
\hat{P}(z,y) &=& \displaystyle \frac{1}{2}\:P(x,x) & {\mbox{if~$y, z \in \{0,x\}$}}\:.
\end{array} \right. \]
Using that the Lagrangian is homogeneous in~$P$ of degree four, we obtain
with the obvious notation~$\hat{A}_{xy} = \hat{P}(x,y)\, \hat{P}(y,x)$ that
\begin{eqnarray*}
{\mathcal{S}}_\mu[\hat{P}] &=& \sum_{y,z \not \in \{0, x\}}
{\mathcal{L}}_\mu[\hat{A}_{y,z}] \:+\: 4 \sum_{y \not \in \{0, x\}}
{\mathcal{L}}_\mu[\hat{A}_{xy}] \:+\: 4\: {\mathcal{L}}_\mu[\hat{A}_{xx}] \\
&=& \sum_{y,z \not \in \{0, x\}}
{\mathcal{L}}_\mu[A_{y,z}] \:+\: \sum_{y \not \in \{0, x\}}
{\mathcal{L}}_\mu[A_{xy}] \:+\: \frac{1}{4}\: {\mathcal{L}}_\mu[A_{xx}]
\end{eqnarray*}
and thus
\[ {\mathcal{S}}_\mu[P] - {\mathcal{S}}_\mu[\hat{P}] \;=\;
\sum_{y \not \in \{0, x\}} {\mathcal{L}}_\mu[A_{xy}] \:+\: \frac{3}{4}\:
{\mathcal{L}}_\mu[A_{xx}] \;\geq\; \frac{3}{4} \sum_{y \in M} {\mathcal{L}}_\mu[A_{xy}]\:. \]
Now we can put in~(\ref{Lbound}) to obtain the inequality
\[ {\mathcal{S}}_\mu[\hat{P}] \;\leq\; \left(1 - \frac{3}{4m} \right) {\mathcal{S}}_\mu[P]\:. \]

Consider a minimal sequence~$P_k \in {\mathcal{P}}^f(H)$. Then
\[ I(m+1, f) \;\leq\; {\mathcal{S}}_\mu[\hat{P}_k] \;\leq\; \left(1 - \frac{3}{4m} \right) {\mathcal{S}}_\mu[P_k]
\;\stackrel{k \to \infty}{\longrightarrow}\; \left(1 - \frac{3}{4m} \right) I(m,f)\:, \]
proving the left inequality in~(\ref{IJin}). Similarly, if we let~$P_k$ be a minimal
sequence of projectors, then the corresponding operators~$\hat{P}_k$ are also projectors
(because~(\ref{hPdef}) is a unitary transformation), and we obtain the right
inequality in~(\ref{IJin}).
\QED

\begin{Prp} \label{prpmin}
Let~$P_k \in {\mathcal{P}}^f$ be a minimal sequence for
the action~(\ref{Sdef}), i.e.
\beq \label{minimal}
\lim_{k \to \infty} {\mathcal{S}}_\mu[P_k] \;=\; I(f,m)\:.
\eeq
Then there is~$\delta>0$ such that
\[ \Tr(E_x P_k) \;\geq\; \delta \spc \forall \:k \in \N,\: x \in M\:. \]
\end{Prp}
{\Proof} We argue by contradiction. Assume that there is~$x \in M$
and a subsequence of~$(P_k)$ (again denoted by~$(P_k)_{k \in \N}$) such that~$\lim_{k \to \infty}
\Tr(E_x P_k) \leq 0$. Then we must clearly
have more than one space-time point, because otherwise~$\Tr(E_x P)=\Tr(P)=f>0$.
We introduce the projector $F =\1-E_x$ and define for large~$k$ the series
of operators~$Q_k$ by
\beq \label{Qkdef}
Q_k \;=\; c_k \: F P_k F
\spc {\mbox{with}} \spc
c_k \;:=\; \frac{f}{\Tr(F P_k)} \:.
\eeq
Since~$\Tr(F P_k) = \Tr(P_k) - \Tr(E_x P_k) \to f$, we know that
\beq \label{lim1}
\lim_{k \to \infty} c_k \;\leq\; 1\:.
\eeq
According to Lemma~\ref{lemmares} {\bf{(i)}}, the operators~$(-Q_k)$ are
positive, and we normalized them such that~$\Tr \,Q_k = f$.
Therefore, the operators~$Q_k$ are again of class~${\mathcal{P}}^f$.
Since they vanish identically on~$E_x(H)$, we can regard them as operators in a discrete
space-time consisting of~$m-1$ space-time points, and thus
\beq \label{lim2}
{\mathcal{S}}_\mu[Q_k] \;\geq\; I(m-1,f)\:.
\eeq
Using that the Lagrangian is homogeneous of degree four, we obtain furthermore
\beq \label{lim3}
{\mathcal{S}}_\mu[Q_k] \;=\; c_k^4\: {\mathcal{S}}_\mu[F P_k F]
\;\leq\; c_k^4\: {\mathcal{S}}_\mu[P_k] \:,
\eeq
where in the last step we used that the Lagrangians of~$P_k$ and~$FP_kF$
coincide away of the space-time point~$x$; more precisely,
\[ {\mathcal{S}}_\mu[P_k] - {\mathcal{S}}_\mu[F P_k F] \;=\;
{\mathcal{L}}_\mu[A_{xx}] + 2 \sum_{y \neq x} {\mathcal{L}}_\mu[A_{xy}]
\;\geq\; 0\:. \]

Taking in~(\ref{lim3}) the limit~$k \to \infty$ and
using~(\ref{lim1}, \ref{minimal}), we obtain in view of~(\ref{lim2}) that
\[ I(m-1,f) \;\leq\; \lim_{k \to \infty}{\mathcal{S}}_\mu[Q_k]
\;\leq\; \lim_{k \to \infty}{\mathcal{S}}_\mu[P_k] \;=\; I(m,f)\:, \]
in contradiction to Lemma~\ref{lemmamon}.
\QED
We point out that, unfortunately, the above argument does not apply to a
minimal sequence of projectors, because the property to be idempotent gets lost
when the operators are restricted similar to~(\ref{Qkdef}) to a subspace of~$H$.

\section{A General Existence Theorem} \label{sec5}
In this section we will show that all the results stated in Section~\ref{sec2}
are a consequence of the following general existence theorem.

\begin{Thm} \label{thm2}
Suppose that~$(P_k)_{k \in \N}$ is a sequence of operators in~$H$
such that the operators $(-P_k)$ are all positive.
Assume furthermore that the corresponding sequence of critical
actions~${\mathcal{S}}[P_k]$ is bounded and that one of the
following two conditions is satisfied:
\begin{description}
\item[(C1)] The local trace is bounded away from zero in the sense that
for suitable~$\delta>0$,
\[ |\Tr(E_x P_k)| \;\geq\; \delta \spc \forall \:k \in \N,\: x \in M\:. \]
\item[(C2)] The spectral weights~$|(A_k)_{xx}|$ are bounded from above in the
sense that for suitable~$C>0$,
\[ |(A_k)_{xx}| \;\leq\; C \spc \forall \:k \in \N,\: x \in M\:. \]
\end{description}
Then there is a subsequence~$(P_{k_l})$ and a sequence of gauge
transformations~$U_l \in {\mathcal{G}}$ such that the gauge-transformed operators
have a limit
\[ P \;:=\; \lim_{l \to \infty} U_l P_{k_l} U_l^{-1} \:. \]
\end{Thm}
The proof of this theorem will be given in Sections~\ref{sec7} and~\ref{sec8}.
Here we simply assume that Theorem~\ref{thm2} holds and deduce the theorems
in Section~\ref{sec2}:

We let~$(P_k)_{k \in \sN}$ be a minimal sequence.
Since all the matrix functionals considered here are continuous,
the limit~$P$ constructed with the above theorem will
certainly be a minimizer. Furthermore, the limit of projectors of rank~$f$
is again a projector of rank~$f$, whereas for general operators the rank only decreases
in the limit. For this reason, it is obvious that by taking limits
we do not leave class of operators under consideration.

When considering the variational principle~(\ref{vary}, \ref{constraint}), the
inequality~$|A_{xx}|^2 \leq \kappa$ shows that condition~{\bf{(C2)}} holds.
Furthermore, ${\mathcal{L}}[P_k] \leq \sum_{x,y} |A_{xy}^2|$. Hence Theorem~\ref{thm2}
applies and gives the desired minimizer~$P$. This proves Theorem~\ref{thmn1},
Theorem~\ref{thm0} and the existence part of Theorem~\ref{thmn5}.
In order to derive the relation~(\ref{range}),
we consider the variation~$P(\tau) = (1+\tau)\:P$. Using that the action is
homogeneous in~$P$ of degree 4, we find that
\[ 0 \;=\; \left.\! \frac{d}{d\tau} {\mathcal{S}}_\mu(P(\tau)) \right|_{\tau=0}
\;=\; 4 \:{\mathcal{S}}_\mu(P) \:, \]
and so the action vanishes.

To prove Theorem~\ref{thmn2}, we decompose the Lagrangian as in~(\ref{dec})
into a sum of two positive terms. This shows that condition~{\bf{(C2)}} is
satisfied, and Theorem~\ref{thm2} applies. In the setting of
Theorem~\ref{thmn3}, the assumption
on the local trace ensures that condition~{\bf{(C1)}} holds, and again Theorem~\ref{thm2}
applies. To prove Theorem~\ref{corres} we let~$P$ be a homogeneous fermionic projector. Then, with~$\sigma$ and~$U$ as in Definition~\ref{defhomo},
\[ \Tr(E_{x_1} P) \;=\; \Tr(P(x_1,x_1)) \;=\;
\Tr(U\: P(x_0, x_0)\: U^{-1}) \;=\;
\Tr(P(x_0, x_0)) \:. \]
Thus the local trace is the same at all space-time points, and
from~(\ref{loctr}) we conclude that
\[ \Tr(E_x P) \;=\; \frac{f}{m} \spc \forall\: x \in M\:. \]
Hence the condition~{\bf{(C1)}} is satisfied, and we can again apply
Theorem~\ref{thm2}.

Finally, to prove Theorem~\ref{thm1}, we can in the case~$\mu < \frac{1}{2n}$
again use the decomposition~(\ref{dec}), whereas in the
critical case~$\mu=\frac{1}{2n}$ Proposition~\ref{prpmin} ensures
that condition~{\bf{(C1)}} holds.
This concludes the proof of all the theorems in Section~\ref{sec2},
provided that Theorem~\ref{thm2} is true.

\section{Gauge Fixing, Rescaling} \label{sec7}
We now enter the proof of Theorem~\ref{thm2}. Thus let~$(P_k)_{k \in \N}$ be a
sequence of operators satisfying the assumptions of Theorem~\ref{thm2}.
We again choose a basis in~$H$ and let~$(.|.)$ be the canonical
scalar product on~$\C^{2nm}$. We let~$\|.\|$ be the corresponding Hilbert-Schmidt
norm, $\|A\| := (\Tr(A^\dagger A))^\frac{1}{2}$.

Our first task is to treat the non-compact gauge group~${\mathcal{G}}$ (as defined
after~(\ref{local})). We denote the equivalence class of gauge-equivalent operators
by~$\langle . \rangle_{\mathcal{G}}$, i.e.
\[ \langle P \rangle_{\mathcal{G}} \;=\; \{ UPU^{-1} \:|\: U \in {\mathcal{G}} \}\:. \]
We consider for any fixed~$k \in \N$ the variational principle
\beq \label{varp}
{\mbox{minimize}} \;\; \left\{ \|Q\| \;|\; Q \in \langle P_k \rangle_{\mathcal{G}} \right\} .
\eeq
If~$(Q_l)_{l \in \N}$ is a minimal sequence of this variational principle,
the Hilbert-Schmidt norms of the~$Q_l$ are uniformly bounded. Thus we can use
a compactness argument to select a convergent subsequence. We conclude that the
variational principle~(\ref{varp}) attains its minimum. We choose for each~$k$ a
minimizer and denote it by~$\hat{P}_k$. We point out that the above construction of
the~$\hat{P}_k$ involves the norm~$\|.\|$ and thus depends on the choice of
our basis of~$H$. This will be no problem in what follows because the
minimizers obtained by choosing different norms will be gauge equivalent.
We refer to our method of arbitrarily choosing
one representative of each gauge equivalence class as {\em{gauge fixing}}; it
can be understood in analogy to the gauge fixing used in electrodynamics
or in general relativity.

In the case that the sequence of operators~$(\hat{P}_k)$ has a subsequence of
bounded Hilbert-Schmidt norm, we can by compactness choose a subsequence
which converges to an operator~$P$. Thus it remains to consider the case when
the Hilbert-Schmidt norm is unbounded for any subsequence of~$(\hat{P}_k)$; in
other words, that
\beq \label{case2}
\|\hat{P}_k\| \;\to\; \infty\:.
\eeq
We introduce new operators~$R_k$ by rescaling the~$\hat{P}_k$,
\[ R_k \;=\; \alpha_k\: \hat{P}_k \spc {\mbox{with}} \spc
\alpha_k \;:=\; \frac{1}{\|\hat{P}_k\|}
\;\stackrel{k \to \infty}{\longrightarrow}\, 0\:. \]
Then obviously~$\|R_k\| \equiv 1$, and thus we can, again after choosing a subsequence,
assume that the~$R_k$ converge,
\[ R_k \;\to\; R\:. \]

It is clear from their construction that the operators~$R_k$ and~$R$ have the
following properties: The operators~$(-R_k)$ and~$(-R)$ are positive and normalized by
\beq \label{p0}
\|R_k\| \;=\; 1 \;=\; \|R\|\:.
\eeq
Their action is computed to be
\beq \label{p2}
{\mathcal{S}}[R_k] \;=\; \alpha_k^4\: {\mathcal{S}}[P_k]\:,\spc {\mathcal{S}}[R] \;=\; 0\:.
\eeq
Finally, the conditions~{\bf{(C1)}} and~{\bf{(C2)}} give
\beq \label{p3}
\left\{ \begin{array}{cl}
|R(x,x)^2| \;=\; 0 \quad \forall\, x \in M &\qquad {\mbox{in case {\bf{(C1)}}}} \\[.5em]
|\Tr(E_x R_k)| \;\geq\; \delta \:\alpha_k \quad \forall \,k \in \N,\: x \in M
&\qquad {\mbox{in case {\bf{(C2)}}}} \:. \end{array} \right.
\eeq

\section{Existence of Minimizers} \label{sec8}
Our goal is to show that the properties~(\ref{p0}--\ref{p3}) contradict the fact that
the~$\hat{P}_k$ are minimizers of~(\ref{varp}) (this then implies that the
case~(\ref{case2}) cannot occur, completing the proof of Theorem~\ref{thm2}).
For any~$x \in M$, the operator~$T:=-R(x,x)$ is positive according
to Lemma~\ref{lemmares}~{\bf{(i)}}. From Lemma~\ref{lemmaspec} we conclude
that the zeros~$(\nu_j)_{j=1,\ldots,2n}$ of its characteristic polynomial
are all real and ordered as in~(\ref{morder}).
Since~${\mathcal{L}}[T^2]=0$, the absolute values of the~$\nu_j$ must all be equal,
and thus there is a parameter~$\nu \geq 0$ such that
\[ \nu_1 = \ldots = \nu_n = -\nu \spc {\mbox{and}} \spc
\nu_{n+1} = \ldots = \nu_{2n} = \nu \:. \]

Let us rule out the case~$\nu>0$. If the condition~{\bf{(C1)}} is satisfied,
we obtain from~(\ref{p3}) that~$0=|T^2| = 2 n \nu^2$, a contradiction.
If on the other hand the condition~{\bf{(C2)}} holds, we know by the
continuity of the spectrum that for large~$k$,
\[ \inf \sigma \!\left((T_k)^2 \big|_{E_x(H)} \right) \;\geq\;
\frac{\nu^2}{2} \]
(with~$T_k:=-R_k(x,x)$). Combining the lower bound~(\ref{Llb2})
with~(\ref{p3}) and~(\ref{p2}), we obtain
\[ \frac{1}{4n}\: \frac{\nu^2}{2}\: \delta^2\, \alpha_k^2 \;\leq\;
{\mathcal{L}}[T_k^2] \;\leq\; {\mathcal{S}}[R_k] \;=\; \alpha_k^4\: {\mathcal{S}}[P_k]\:. \]
Dividing by $\alpha_k^2$ and taking the limit~$k \to \infty$, we obtain a contradiction to the boundedness of the sequence~${\mathcal{S}}[P_k]$.

It remains to consider the case~$\nu=0$ where the operator~$T$ is nilpotent.
As in the proof of Lemma~\ref{lemmaspec}, we approximate~$T$ by the
strictly positive operators~$T_\varepsilon = T + \varepsilon S$.
Diagonalizing the~$T_\varepsilon$ by unitary
transformations~$U_\varepsilon$ on~$E_x(H)$, the diagonal matrices~$U_\varepsilon T_\varepsilon
U_\varepsilon^{-1}$ converge to zero as~$\varepsilon \to 0$.
Hence for any~$\Psi \in H$,
\[ \bra \Psi \:|\: U_\varepsilon \,T\, U_\varepsilon^{-1} \, \Psi \ket \:+\:
\varepsilon\, \bra \Psi \:|\: U_\varepsilon \,S\, U_\varepsilon^{-1} \, \Psi \ket \;=\;
\bra \Psi \:|\: U_\varepsilon \,T_\varepsilon\, U_\varepsilon^{-1} \, \Psi \ket
\;\stackrel{\varepsilon \to 0}{\longrightarrow}\; 0 \:. \]
Since the summands on the left are both positive, we conclude
that~$\bra \Psi \:|\: U_\varepsilon T U_\varepsilon^{-1} \, \Psi \ket \to 0$
for all~$\Psi \in H$ and thus
\[ \lim_{\varepsilon \to 0} U_\varepsilon T U_\varepsilon^{-1} \;=\; 0\:. \]
For given~$\kappa>0$ we choose~$\varepsilon$ such that
$\|U_\varepsilon T U_\varepsilon^{-1} \| < \kappa/2$ and subsequently~$k$ so large that
$\|T_k - T\| < \kappa/(2\, \|U_\varepsilon\|\, \|U_\varepsilon^{-1}\|)$. Then
\[ \|U_\varepsilon \,T_k\, U_\varepsilon^{-1}\|
\;\leq\; \|U_\varepsilon\|\, \|T_k-T\|\, \|U_\varepsilon^{-1}\|
\:+\: \|U_\varepsilon \,T \,U_\varepsilon^{-1}\| \;\leq\; \kappa \:. \]
Since~$\kappa$ can be chosen arbitrarily small, we conclude that
there is a subsequence of the~$T_k$ (which we denote again by~$(T_k)_{k \in \N}$)
together with unitary transformations~$U_k$ such that
\[ \lim_{k \to \infty} U_k\,T_k\,U_k^{-1} \;=\; 0\:. \]
Extending the~$U_k$ by the identity to the subspaces~$E_y(H)$, $y \neq x$,
we obtain a sequence of gauge transformations such that
\[ U_k \,R_k\, U_k^{-1} \;\to \; \tilde{R}\:. \]
Since these gauge transformations act only on~$E_x(H)$, it is clear
that~$R(y,z) = \tilde{R}(y,z)$ if~$y,z \neq x$. By construction,
$\tilde{R}(x,x)=0$. The Schwarz inequality, Lemma~\ref{lemmares}~{\bf{(ii)}},
tells us that also the entries~$\tilde{R}(x,y)$ and~$\tilde{R}(y,x)$ for~$y \neq x$ vanish.
Since we chose the operators~$\hat{P}_k$ such that their Hilbert-Schmidt
norm was minimal among all gauge-equivalent operators,
the Hilbert-Schmidt norm of the operators~$R_k$ (which were obtained from the~$\hat{P}_k$
only by rescaling) cannot be decreased by a subsequent gauge
transformation, and thus~$\|U_k \,R_k\, U_k^{-1}\| \geq \|R_k\|$.
Taking the limit~$k \to \infty$, we find that~$\|\tilde{R}\| \geq \|R\|$.
Since these operators coincide up to matrix elements where~$\tilde{R}$ vanishes,
the operators~$\tilde{R}$ and~$R$ must coincide.
In particular, $R(x,x)=0$.

We conclude that the diagonal entries~$R(x,x)$ of~$R$ all vanish. Again
applying the Schwarz inequality, Lemma~\ref{lemmares}~{\bf{(ii)}}, we see that
the off-diagonal entries of~$R$ are also zero. Thus~$R=0$,
in contradiction to~(\ref{p0}). \\

\noindent
{\em{Acknowledgments:}} I would like to thank Niky Kamran and
Daniela Schiefeneder for helpful
comments on the manuscript. I am grateful to the Erwin Schr\"odinger Institute,
Wien, for its hospitality while I was working on this paper.

\noindent
NWF I -- Mathematik,
Universit{\"a}t Regensburg, 93040 Regensburg, Germany, \\
{\tt{Felix.Finster@mathematik.uni-regensburg.de}}


\addcontentsline{toc}{section}{References}
\begin{thebibliography}{99}
\bibitem{Bognar} J.\ Bognar, ``Indefinite Inner Product Spaces,''
Ergebnisse der Mathematik und ihrer Grenzgebiete, Band~{\bf{78}},
{\em{Springer Verlag}}, New York - Heidelberg (1974)
\bibitem{PFP} F.\ Finster, ``The Principle of the Fermionic Projector,''
{\em{AMS/IP Studies in Advanced Mathematics}} {\bf{35}} (2006)
\bibitem{GLR} I.\ Gohberg, P.\ Lancaster, L.\ Rodman, ``Matrices and
Indefinite Scalar Products,'' {\em{Birkh{\"a}user Verlag}} (1983)
\end{thebibliography}
\end{document}